# Electronic Hybridization of Large-Area Stacked Graphene Films

Jeremy T. Robinson[1], Scott W. Schmucker[1], C. Bogdan Diaconescu[2], James P. Long[1], James C. Culbertson[1], Taisuke Ohta[2], Adam L. Friedman[1], Thomas E. Beechem[2]

[1]*Naval Research Laboratory, Washington, DC 20007, USA*
[2]*Sandia National Laboratories, Albuquerque, NM 87185, USA*

**Abstract**

Direct, tunable coupling between individually assembled graphene layers is a next step towards designer two-dimensional (2D) crystal systems, with relevance for fundamental studies and technological applications. Here we describe the fabrication and characterization of large-area (> cm$^2$), coupled bilayer graphene on SiO$_2$/Si substrates. Stacking two graphene films leads to direct electronic interactions between layers, where the resulting film properties are determined by the local twist angle. Polycrystalline bilayer films have a "stained-glass window" appearance explained by the emergence of a narrow absorption band in the visible spectrum that depends on twist angle. Direct measurement of layer orientation *via* electron diffraction, together with Raman and optical spectroscopy, confirms the persistence of clean interfaces over large areas. Finally, we demonstrate that interlayer coupling can be reversibly turned off through chemical modification, enabling optical-based chemical detection schemes. Together, these results suggest that individual 2D crystals can be individually assembled to form electronically coupled systems suitable for large-scale applications.

The discovery and development of graphene [1, 2] has drawn attention to a class of materials broadly defined as two-dimensional (2D) crystals.[3] As the field matures in understanding and exploiting single 2D crystals, new research is emerging to build up materials from individual layers forming "designer" thin films.[4-7] While simple material combinations may be realized through direct growth,[8, 9] many others will require the physical stacking of individual layers. However, the top-down synthesis of multilayer structures is often plagued by surface contamination, which limits direct van der Waals contact between layers.[10, 11] In the simplest case— the stacking of two graphene layers— new properties arise when clean interfaces are realized.[12-14] It is already clear that bilayer graphene is a diverse material system because the film properties vary as a function of relative orientation of one layer to the other. Such twisted graphene systems,[15] whose theoretical [16-18] and experimental [11-14, 19-22] properties are now being intensely considered, will guide our development of other synthetic van der Waals films.

In this work we describe the top-down synthesis and characterization of coupled bilayer graphene films that exhibit intimate, clean contact over macroscopic areas (>cm$^2$). The observation of widespread "colored" domains in these stacked films reveals that coupling between layers results in new properties not intrinsic to the individual components, an effect not observed previously in large-area artificial bilayer films on SiO$_2$/Si substrates. To determine if these colored domains are an intrinsic feature of the film we use Raman spectroscopy, which has proven to be useful in quantifying various degrees of interaction between graphene layers.[11, 13, 19, 23, 24] Interlayer hybridization in twisted bilayer graphene (TBG) results in measurable changes in the intensity, position, and shape of the characteristic G (~1600 cm$^{-1}$) and 2D (~2700 cm$^{-1}$) Raman peaks; as such, a strong one-to-one correlation between Raman intensity and twist angle has been established.[11, 13] In addition to changes in the Raman response, the hybridization of the Dirac cones in TBG [21] results in changes of the optical conductivity, including the emergence of an absorption peak [11, 20] in the relatively wavelength-independent spectrum of single-layer or Bernal bilayer graphene.[25, 26] In this work, optical spectroscopy reveals the emergence of an adsorption peak in the visible spectrum for TBG domains with twist angles between ~10-16°, which we have independently confirmed using low energy electron diffraction. Analogous to the rapid development of graphene research provided by simply "visualizing" graphene,[27, 28] the visualization of specific twist orientations with an optical microscope should further enable the rapid study of TBG systems. As such, we use optical microscopy to demonstrate that interlayer coupling can be effectively switched "on" and "off" through chemical functionalization of the top surface.

Graphene films were grown *via* low-pressure chemical vapor deposition (CVD) in Cu foil enclosures [29] and transferred onto SiO$_2$ (100nm)/Si substrates using conventional wet-chemistry techniques,[30-32] including the use of a Poly(methyl methacrylate) (PMMA) protective layer [33] and APS Copper Etchant 100 (Transene Company, Inc.). Before etching Cu, we oxygenate the Transene etchant, which we find results in fewer carbonaceous residues after PMMA removal. Immediately after transferring the PMMA/graphene film from a H$_2$O bath the substrate is spun at 2000 rpm, then 4000 rpms to remove bulk water (Fig. 1A). The sample is subsequently heated on a hot plate (T=85°C, 3min; T=150°C, 20 min.) and then submerged in an acetone bath, rinsed in acetone and IPA and dried with $N_2$. Bilayer samples are generated by sequentially transferring a second CVD graphene layer onto the graphene/$SiO_2$/Si substrate and repeating the steps described above. The polycrystalline nature of the starting CVD graphene results in bilayers with regions of varying twist angle and is referred to here as twisted bilayer graphene.

**Results and Discussion**

Optical inspection of as-fabricated TBG samples reveals expected features such as wrinkles and folds,[34, 35] together with an unexpected patchwork of colored domains that vary in size and shape across the sample. Figure 1B shows an optical microscope image (Olympus DP25 CCD camera) of a TBG film with distinct regions that appear "red", "yellow", and "blue". No filters are necessary to see these colored features (see Supporting Information). Notably, subsequent thermal annealing (*e.g.*, 400°C in Ar/$H_2$)[36] or additional solvent cleaning does not change the extent of these colored regions, suggesting their presence is not related to extrinsic processing residues. Figure 1C shows a higher resolution image of a different TBG sample in which the second (top) graphene layer is non-continuous. Here we find the colored domains are only observed in the bilayer regions and not the neighboring single-layer regions. In addition to these macroscopic colored features, small isolated islands are observed (bluish "dots" Fig. 1C; discussed below), together with the wrinkles/folds mentioned earlier.

Atomic force microscope (AFM) imaging at the boundary between a bilayer and single-layer region shows an increased level of roughness in the bilayer region (Fig. 1E). This increased roughness is in the form of small isolated islands that have typical footprints of less than 1 $\mu m^2$ and height of ~10-40 nm and are separated by relativity smooth areas. Similar islands have been observed elsewhere when multiple graphene layers are individually assembled,[10, 13] where the largest islands align well with the bluish "dots" observed optically (Fig. 1C). Cross-sectional imagining of stacked graphene layers confirms the presence of locally trapped amorphous material in such samples, separated by pristine regions in direct van der Waals contact.[10] The ability of trapped (inter-layer) hydrocarbons/ adsorbates to segregate over micron-scale distances gives rise to these isolated islands.[10] We note that using a similar sample fabrication procedure as described above, we have measured coupling between atomically flat epitaxial graphene on SiC and transferred CVD graphene films,[21, 37] supporting the presence of clean interface regions between such isolated islands.

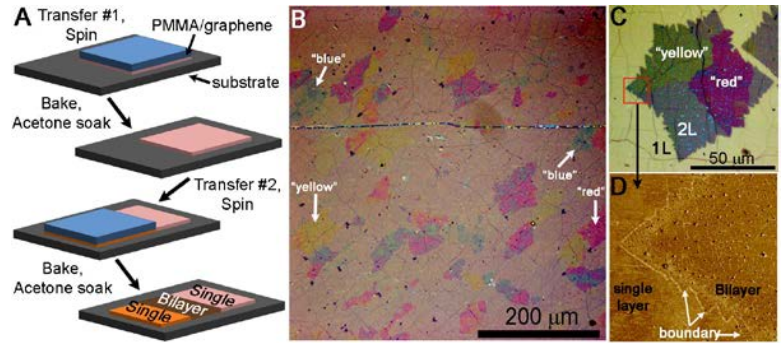

**Figure 1:** Fabrication and characterization of two stacked graphene films. **(A)** Schematic of the process flow used in forming bilayer graphene from single-layer CVD graphene films. **(B)** Optical microscope image of a TBG film. The "red", "yellow", and "blue" domains are labeled. **(C)** Sample in which the top (second) graphene layer is only a partial layer, forming single layer (1L) and bilayer (2L) regions. **(D)** AFM phase image (15 × 15 μm$^2$) showing the boundary between the single layer and bilayer region highlighted by the red box in (C).

The direct correlation between the Raman spectral response and layer orientation in TBG films[11, 13] provides a rapid means to determine if interlayer coupling is present, as well as to coarsely identify twist angle in coupled bilayers. Arguably the most dramatic feature in TBG Raman spectra is a substantial enhancement (>20×) in the G peak intensity when the laser photon energy ($E_{photon}$) is approximately equal to the energy level where the Dirac cones overlap/ hybridize. Because the position of band overlap is dependent on twist angle ($\theta$), the orientation at which this maximum Raman enhancement occurs is dubbed the critical angle ($\theta_c$), with a critical energy $E_{\theta c}$. Conveniently, this G peak enhancement occurs in a somewhat narrow range— when $\theta$ is within a few degrees (about ± 2°) of $\theta_c$ or when $E_{photon}$ is within a few hundred meV of $E_{\theta c}$. When outside of this window, the Raman spectra can only indicate if $\theta < \theta_c$ or $\theta > \theta_c$.[11, 13]

Unlike two non-interacting stacked graphene sheets, the Raman spectra for our TBG samples show a rich variation in peak intensities and shapes as would be expected for two interacting layers with distinct twist angles. Figure 2 shows Raman maps and spectra from TBG films in Fig. 1, which contain red, yellow, and blue domains. Figures 2A and B show a map of the G/2D peak ratio (integrated intensity) for the TBG film in Fig. 1C, measured at two different wavelengths (488nm and 532nm). Close inspection of the map reveals six unique intensity ratios in the TBG region (labeled). Notably, the colored domains in the optical image align very well with the domains imaged by Raman. This result, together with many other Raman/ optical image comparisons (not shown), provides strong support for the colored domains being related to an intrinsic property of the film and being uniquely dependent on the twist angle between the graphene layers.

Individual Raman spectra (Fig. 2C,D) from the samples shown in Fig. 1B and C can be understood in the frameworks recently presented by Havener *et al.*[11] and Kim *et al.*.[13] The use of two different photon energies allows us to order the spectra in terms of increasing twist angle through analyzing the

G/2D peak ratios and peak positions. A strong G peak enhancement occurs at the yellow domain using $E_{photon}$= 2.54 eV (488 nm) and at the red domain using $E_{photon}$= 2.33 eV (532 nm). If we assume $E_{photon}$ here is equal to $E_{\theta c}$, this corresponds to $\theta_{yellow} \approx 14.5°$ and $\theta_{red} \approx 12.5°$.[11] Comparing histograms of the normalized integrated G peak intensity ($A_G$) for many spectra within these same regions sheds light on the closeness of $E_{photon}$ to $E_{\theta c}$, as well as on the angular distribution or deviation in $\theta$ (Fig. 2E,F). For example, at the resonance condition ($E_{photon}= E_{\theta c}$) for $\theta_c= 12.5°$ the intensity of $A_G$ decreases by 50% within approximately ± 1° and by 80% within approximately ± 2.5° twist variation.[11] If the twist angle of the red domain equals 12.5° with a small twist deviation (± 1°), the histogram at $\lambda$= 532 nm should appear like that in Fig. 2E, having a high mean value and left skew. On the other hand, if $\theta$ ($E_{photon}$) differs by even 1° (150 meV) from $\theta_c$ ($E_{\theta c}$), then the $A_G$ intensity distribution will have a lower mean value (by up to ~50%) and have a more symmetric or right skew as observed with the yellow domain for $\lambda$= 488 nm (Fig. 2F). Together, this implies $\theta_{red} \approx 12.5°± 1°$ and that $\theta_{yellow}$ is slightly larger than 14.5°. Finally, for the blue domain, the slight increase in the G/2D ratio and upshift of the 2D peak from $\lambda$ = 488 nm to $\lambda$ = 532 nm indicates it is by several degrees smaller than $\theta_{red}$.

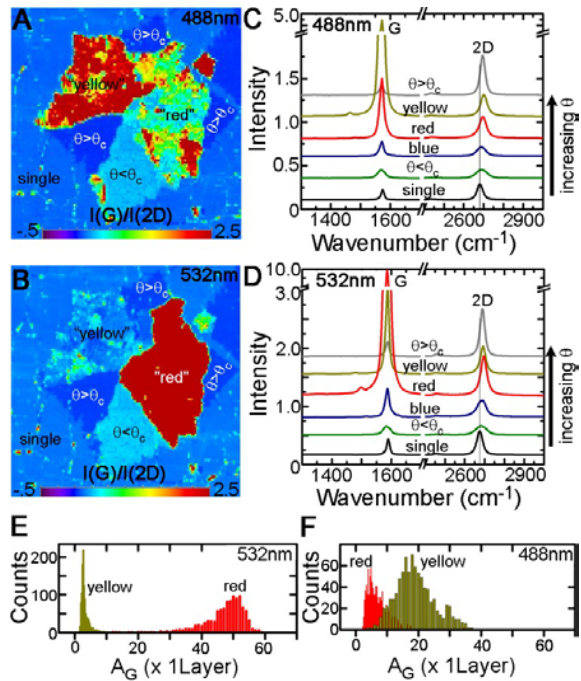

**Figure 2:** Raman mapping and spectra of TBG domains using a **(A,C)** 488nm (2.54eV) and **(B,D)** 532nm (2.33eV) laser. The map is of the TBG region shown in Fig. 1C. **(C, D)** Averaged spectra from different regions of the sample as labeled. The same areas are analyzed for both wavelengths. The spectra are offset for clarity and are arranged in order of increasing twist angle. The G and 2D peaks are labeled. **(E,F)** Histogram of the normalized integrated G peak intensity ($A_G$) taken from individual spectra within the "yellow" and "red" regions shown in (A) and (B) at each wavelength (labeled). **(E)** For the "red" distribution: mean = 46.7, median = 48.1, range = 46.2. **(F)** For the "yellow" distribution: mean = 18.6, median = 17.8, range = 37.3.

The presence of colored TBG domains suggests there are unique differences in the absorption spectrum that are dependent on twist angle. The electronic hybridization in TBG leads to changes in the optical transition matrix elements and joint density of states (JDOS), which can strongly affect the absorption properties.[11, 20] Figure 3 shows the measured and calculated optical contrast spectra for a red, yellow, blue, and large-angle bilayer domain ($\theta > \theta_c$ for $E_{photon}$= 2.54 eV), together with a single-layer region. The contrast spectrum is defined as $C(\lambda) = [R_o(\lambda) - R(\lambda)]/R_o(\lambda)$, where $R_o(\lambda)$ is the reflection spectra from the substrate (SiO$_2$/Si) and $R(\lambda)$ is reflection from the film plus substrate [28, 38] (see Supporting Information). For single-layer graphene we find a reasonably good match between the measured and calculated $C(\lambda)$ using an index of refraction for graphene of $n_g$=2.6-1.3$i$ [28] and thickness of 0.34 nm, together with literature values of $n(\lambda)$ for SiO$_2$ and Si.[39] For two-layer graphene, the measured $C(\lambda)$ (TBG domains) and calculated $C(\lambda)$ (Bernal stacking) do not agree well. In particular, $C(\lambda)$ for the colored TBG domains is not the smooth envelope shape, but instead there is a shoulder feature unique to each domain.

Since contrast here is approximately proportional to absorption,[11, 20] the difference between two contrast spectra highlights relative absorption features of one domain to another. Hence we define a contrast difference spectrum, $\delta C_i(\lambda) = C_i(\lambda) - C_{\theta > \theta_c}(\lambda)$, where the subscript i indicates yellow, red, or blue. We chose a large-angle domain as a common reference since its properties in visible wavelengths are close to that of two independent graphene layers. Figure 3B compares $\delta C(\lambda)$ among the red, yellow, and blue domains. For each curve, there is a distinct peak ($E_{peak}$), indicating enhanced relative absorption at specific regions of the visible spectrum.

Given the knowledge of the substrate determined above, it should in principle be possible to model the complex dielectric function $\varepsilon(\omega) = \varepsilon_1(\omega) - i\varepsilon_2(\omega)$ of TBG that produces the strong features in the contrast-difference data. ($\varepsilon_2(\omega)$ is directly related to the intrinsic material absorption, and includes contributions from the TBG hybridization states.) To carry out the modeling, we used a commercial software package able to model the reflectance from layered stacks.[40] We were able to obtain rough agreement by using the time-honored approach of modeling narrow absorptions with Gaussian or Lorentzian oscillators, but found the fits improved significantly when we employed a more flexible model that, in a Kramers-Kronig consistent fashion, reproduces the van Hove singularity (vHs) and linear density of states expected for the TBG absorption.[11, 41] Our fits to $\delta C(\lambda)$ for each domain angle are shown with dashed/dotted lines in Fig. 3B, and the corresponding $\varepsilon_2(\lambda)$ are included in Fig. 3B inset. While qualitative in nature, the fit reproduces the small dip after the main peak, as well as the increasing tail at lower wavelengths, indicating the singularity plus linear shape of the JDOS is a good approximation for the presence and shape of the absorption feature. In addition, the FWHM of the absorption feature is approximately 0.25 eV, close to that predicated by DFT calculations.[11] Assuming here that $E_{peak}$ equals $E_{\theta c}$, this corresponds to $\theta_{blue} \approx 10.5°$, $\theta_{red} \approx 12.5°$, $\theta_{yellow} \approx 15.5°$,[11] in excellent agreement with the Raman results and analysis in Figure 2.

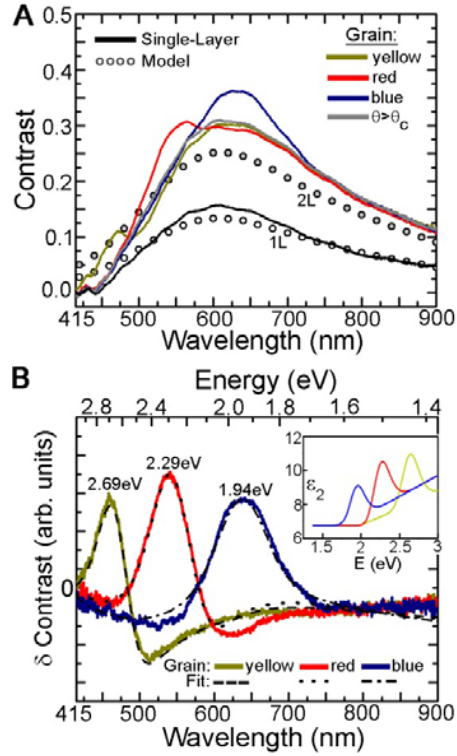

**Figure 3:** Averaged contrast spectra (C($\lambda$)) for TBG domains. **(A)** Measured C($\lambda$) for a red, yellow, blue, and large-angle ($\theta>\theta_c$) domain, together with a single-layer region. All spectra are referenced to the SiO$_2$(100 nm)/Si substrate. The calculated contrast for single-layer (1L) and two-layer (2L) graphene is included. The red, yellow, and $\theta>\theta_c$ domain were measured from Fig. 1C. **(B)** Contrast difference of the yellow, red, and blue domains with respect to the large-angle domain, together with the best-fit curve calculated using $\varepsilon_2(\omega)$ shown in the inset.

To directly quantify interlayer twist angles and variations within one layer's in-plane orientation, we conduct micro-diffraction experiments using low energy electron microscopy/diffraction (LEEM/LEED). Unlike LEED of TBG on SiC,[37] the LEED spots of graphene and TBG on SiO$_2$ are diffuse (Fig. 4A,B) due to the nanometer-scale substrate roughness of SiO$_2$.[42, 43] Since diffracting electrons from the bottom layer in TBG on SiO$_2$ are greatly attenuated in these samples, we presume that the orientation of the diffraction pattern represents primarily that of the top surface layer. As such, we only measure one family of diffraction spots on either TBG or single-layer graphene (Fig. 4A, B).

Mapping diffraction patterns allows direct correlation between the angular orientation and spatial extent of TBG domains. Figure 4C shows such a false color map of LEED angular orientation. The measurement was carried out by translating the partial TBG film (shown in Fig. 4D) with respect to the electron beam (5 μm diameter) over 250 x 200 μm$^2$ (50 × 40 data points). We specify the orientation of the diffraction pattern from the angle defined in Fig. 4A, based on the intensity profile along the dotted arc (Fig. 4B). Distinct angle-specific domains (labeled) in both the single and bilayer regions become apparent as highlighted by the black outlines in Fig. 4C and D. Cross-sectional line scans across TBG and the neighboring single-layer regions allows us to quantify the relative twist angle between the upper and lower layers (Fig. 4E). A histogram of all diffraction spots also reveals the relative LEED orientation, as well as the deviation in graphene's in-plane orientation within a domain. Each peak (or cluster) in Fig. 4F represents a domain within Fig. 4C, and shows a typical domain angular variation of ± 1°. Using this same process to determine twist angles of other colored domains (see Supporting Information), we find $\theta_{blue}$ = 11° ±1°, $\theta_{red}$ = 13° ±1° and $\theta_{yellow}$ = 15° ±1°, in good agreement with Raman and optical spectroscopy measurements despite the large angular uncertainties due to broadened diffraction patterns.

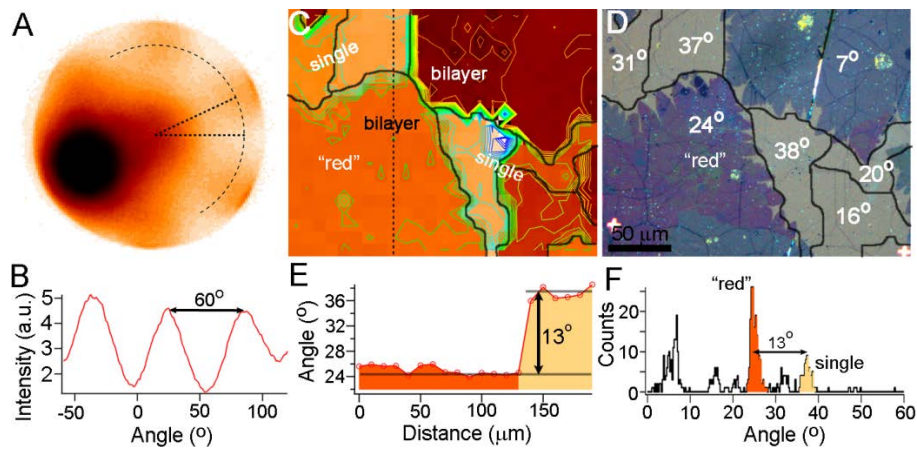

**Figure 4:** LEED analysis of TBG films. **(A)** A typical LEED pattern of a TBG film measured using 5 μm diameter area-selective aperture and an electron energy of 55 eV. **(B)** Intensity profile along the dotted arc in (A). **(C)** A false color map of the LEED angular orientations of a TBG film. **(D)** Optical microscope images of the TBG film measured in (C). The numbers indicate the approximate LEED orientations. **(E)** Vertical line scans taken at the black dash line in (C). **(F)** Histogram from the angular contour plot in (C). The "red" and neighboring single-layer domain from (C) are specified in (F).

The macroscopic appearance of primarily three distinct, colored TBG domains is striking since the energy of the vHs feature continuously varies with twist angle.[12] As such, this may be an indication that the fabrication process leads to preferred TBG orientations. A larger optical microscope survey does show that some TBG domains appear to have distinct, mixed color phases (*e.g.*, Fig. 1B; Fig. 5B). We are currently conducting higher spatial and spectral resolution imaging to determine if there are statistically significant differences in the areal coverage of specific TBG domains. It can be said that the observed mixed colored domains are at least in part due to the measured angular (or twist) deviation within each domain, where a few degree in-plane rotation can result in a 200-300 meV shift in the absorption feature (Fig. 3), as well as $E_{0c}$. As such, it is most common to find mixing of blue ($\theta_{blue}$= 11° ±1°) and red ($\theta_{red}$= 13° ±1°), or red ($\theta_{red}$= 13° ±1°) and yellow ($\theta_{yellow}$= 15° ±1°) domains, but not the blue ($\theta_{blue}$= 11°

±1°) and yellow ($\theta_{yellow}$= 15° ±1°) domains which have an angular separation larger than the angular deviation of domains in these samples. Overall improvements in TBG domain size could be greatly facilitated through the use of highly-orientated graphene films as recently demonstrated on Au-foil substrates,[44] while local twist angle variations may be reduced through further optimizing the transfer process, including the use of smoother starting surfaces and low surface energy solvents during transfer and drying.[45]

Finally, we demonstrate lithographic control of interlayer coupling by selective fluorination of the top graphene layer using XeF$_2$ gas.[46] Following fluorination, even for very short XeF$_2$ exposures (*e.g.*, 1 s, 1 Torr XeF$_2$), color contrast is fully quenched in TBG domains with twist angles between 10-16°. This is seen in Fig. 5A following fluorination in a pattern defined by a PMMA mask. Upon fluorine desorption (T= 175°C for 1hr in flowing Ar) the color contrast returns, demonstrating the sensitivity of interlayer coupling to covalent functionalization and the usefulness of optical characterization as a probe for interlayer hybridization. The reappearance of interlayer coupling after de-fluorination is noteworthy since defects are often introduced during the adsorption/desorption process,[46] highlighting the overall robust nature of coupling. Fluorine adsorption changes the electronic properties of graphene by reducing the charge in the conducting π orbitals, by introducing scattering centers, and by opening band gaps.[46] Fluorine also structurally deforms the graphene skeleton as bond angles shift due to transiting $sp^2$ to $sp^3$ carbon bonding, which is confirmed by the emergence of the D peak in Raman spectroscopy [46, 47] (not shown). Thus, decoupling *via* mechanical separation or *via* quenching of the hybridized electronic state are promising routes for optical-based chemical detection using TBG films with twist angles between 10-16°. In addition to chemical or mechanical routes, it is likely that electrostatic gating will provide tunable control over interlayer coupling and bandstructure, similar to that reported for Bernal (AB-stacked) bilayer graphene. [48, 49]

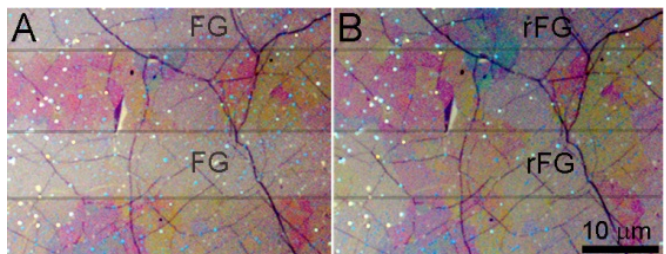

**Figure 5:** The local addition **(A)** and removal **(B)** of fluorine on the top surface of TBG films. Lines help guide the eye where TBG was fluorinated (labeled "FG") and reduced (labeled "rFG").

**Conclusion**

In summary, we show that interlayer coupling between two large-area, individually stacked graphene films is possible and that interlayer coupling enables the direct visualization of specific bilayer orientations on SiO$_2$/Si substrates. The diffusion of interlayer adsorbates into localized islands facilitates the formation of atomically clean regions that can electronically hybridize. The polycrystalline nature of the initial CVD graphene leads to bilayer films whose properties depend on the relative twist angle between layers. The interlayer coupling results in the emergence of a relatively narrow absorption peak in the visible for twist angles between 10-16°, where the twist angle variations within a single domain are found to be approximately ± 1° in these samples. By selectively functionalizing the top surface of these bilayer films, the interlayer coupling is effectively quenched, opening possibilities of optical-based chemical detection applications. We believe the results presented here will facilitate the formation of various large-area homo- and heterogeneous bilayer systems, where interlayer hybridization can result in exploitable electronic or optical properties.

**Methods**

Raman measurements were performed using a confocal geometry. Dichroic beam splitters were used to reflect single-mode 488 nm or 532 nm laser light onto the excitation / detection optical axis. A 100× microscope objective (NA = 0.65) focused the laser (spot ≈0.4 μm) onto the sample and gathered Raman scattered light for detection. The Raman scattered photons were dispersed in a half-meter Acton Sp-2500 spectrometer and were detected using a Princeton Instruments CCD array (Spec-10:400BR back-thinned, deep-depleted array).

Optical spectroscopy measurements were performed with an inverted microscope (Nikon TE2000) coupled directly through a side port to an imaging spectrometer (Princeton Instruments MicroSpec 300). The spectrometer was equipped with an 8μm pixel CCD (Andor Model 885) and entrance slit assembly that could be temporarily moved aside for imaging. Spatial registration between images and spectra was assured because the spectrometer grating (50 groove/mm) could be exchanged under computer control with a mirror for direct imaging. The sample was illuminated through the 20× objective (NA = 0.45) with a quartz-tungsten-halogen lamp. The Koehler illuminator was apertured to restrict the incident rays to be within ±10° of the sample normal. Reflected light was collected by the same objective. For optical modeling of the contrast-difference spectra, to reproduce the expected van Hove singularity and linear density of states in TBG, we used the software's [40] built-in functions psemi0 and psemi1.[40, 50, 51]

To slightly improve the diffraction contrast during LEED measurements, we first deposited a self-assembled monolayer of hexamethyldisilazane (HMDS) on $SiO_2$[52] before TBG deposition, in order to reduce surface roughness. We then acquired the LEED images *via* a 5 μm illumination aperture selecting the single graphene layer and then the TBG domains (see Supporting Information). In order to insure that both single and bilayer regions have uniform orientations in the field of view, dark field LEEM images were acquired as shown in Supplemental Figures S4. Fluorination experiments were carried out using a Xactix® $XeF_2$ etching tool. In Figure 5A the sample was exposed in Pulse Mode with the following parameters: 10 cycles, 30 s/cycle, 1 Torr $XeF_2$, 35 Torr $N_2$.

**Acknowledgements**

The work at the Naval Research Laboratory was supported by the Office of Naval Research and NRL's Nanoscience Institute. J.T.R. is grateful for continued technical support from D. Zapotok and D. St Amand and C.D. Cress for assistance with Graphic Design. This research was performed while S.W.S held a National Research Council Research Associateship Award at the Naval Research Laboratory. The work at Sandia National Laboratories was supported by the U.S. DOE Office of Basic Energy Sciences (BES), Division of Materials Science and Engineering and by Sandia LDRD. Sandia National Laboratories is a multiprogram laboratory managed and operated by Sandia Corporation, a wholly owned subsidiary of Lockheed Martin Corporation, for the U.S. Department of Energy's National Nuclear Security Administration under Contract DE-AC04-94AL85000.

# Supporting Information

**Optical Imaging:**

The color contrast of twisted bilayer graphene (TBG) domains can be increased by optimizing the exposure time of the CCD camera (Olympus DP25) mounted on the optical microscope. Figure S1 shows examples of four different images acquired at different exposures, without the use of any filters or polarizers. Optical images shown in the main text were acquired at exposures times that produced optimal contrast.

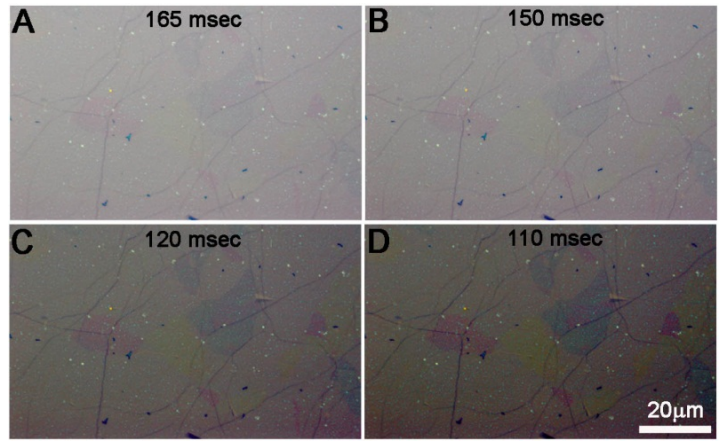

**Figure S1:** Series of optical microscope images acquired at different exposure times as labeled.

**Optical Contrast Modeling:**

Following previous reports[1, 2] for calculating the optical contrast of the trilayer system graphene/SiO$_2$/Si (Figure S2A), it has been shown that contrast is defined by:

$$C(\lambda) = \frac{R_o(\lambda) - R(\lambda)}{R_o(\lambda)}$$

where $R(\lambda)$ is the reflection spectra from graphene/SiO$_2$/Si and $R_o(\lambda)$ is the reflection spectra from SiO$_2$/Si. The reflection spectrum is defined by

$$R(\lambda) = r(\lambda)r^*(\lambda)$$

where

$$r(\lambda) = \frac{r_a}{r_b}$$

$$r_a = r_1 e^{i(\beta_1+\beta_2)} + r_2 e^{-i(\beta_1-\beta_2)} + r_3 e^{-i(\beta_1+\beta_2)} + r_1 r_2 r_3 e^{i(\beta_1-\beta_2)}$$

$$r_b = e^{i(\beta_1+\beta_2)} + r_1 r_2 e^{-i(\beta_1-\beta_2)} + r_1 r_3 e^{-i(\beta_1+\beta_2)} + r_2 r_3 e^{i(\beta_1-\beta_2)}$$

and

$$r_1 = \frac{n_0-n_1}{n_0+n_1}; \quad r_2 = \frac{n_1-n_2}{n_1+n_2}; \quad r_1 = \frac{n_2-n_3}{n_2+n_3}$$

The phase shift (β) due to changes in the optical path is given by: $\beta_1 = 2\pi n_1 \frac{d_1}{\lambda}$; $\beta_2 = 2\pi n_2 \frac{d_2}{\lambda}$. We note that the index of refraction (*n*) for SiO$_2$ and Si is wavelength dependent [3]. Because the graphene refractive index $n_g$ is sensibly constant over our wavelength range [4], we use for convenience the fixed value $n_g$=2.6-1.3i considered by Blake et al. [1]. We use a thickness per layer of graphene of $d_1$=0.34 nm, the target thickness of SiO$_2$ layer ($d_2$= 100 nm), and assume the Si substrate is semi-infinite. To calculate $R_o(\lambda)$, we assume $d_1$ = 0 and $n_0$=$n_1$=1. Figure S2B shows an example of C(λ) calculated for graphene on three different SiO$_2$ thicknesses. The contrast for 90nm thick SiO$_2$ agrees well with Reference [1].

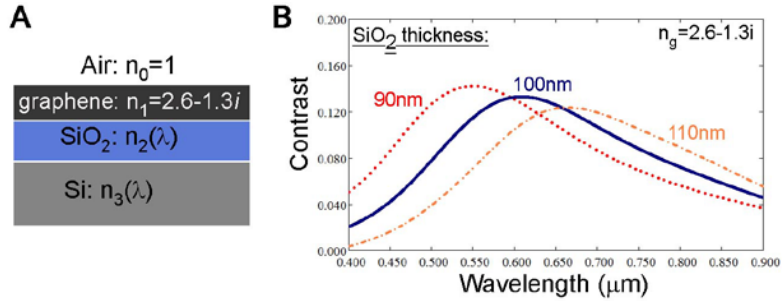

**Figure S2:** (A) Schematic illustrating the optical stack considered here. (B) Comparison of C(λ) for graphene on three different SiO$_2$ layers as labeled: i) 90nm, ii) 100nm, iii) 110nm thick.

**Low Energy Electron Microscopy (LEEM):**

Owing to the contrast of the bilayer regions in both optical microscopy and LEEM, it is possible to investigate locally the same area of the sample with both techniques and thus provide a direct correlation between the colors of the optical domains and the bilayer graphene twist angle as measured via LEED. Figure S3A show three bilayer domains of uniform yellow, red, and blue colors at the edge of the TBG layer. The shape of the bilayer edge is used to locate the same bilayer domains in LEEM as shown in Figure S3B. We then acquired the LEED images via a 5 μm illumination aperture selecting the single graphene layer (Fi. S3D) and then the TBG domains (Fig. S3E) and measured the corresponding twist angles of $\theta_{blue}$ = 11°±1°, $\theta_{red}$ = 13°±1°, and $\theta_{yellow}$ = 15°±1°. In order to insure that both single and bilayer regions have uniform orientations in the field of view, dark field LEEM images were acquired as shown in Fig. S3C.

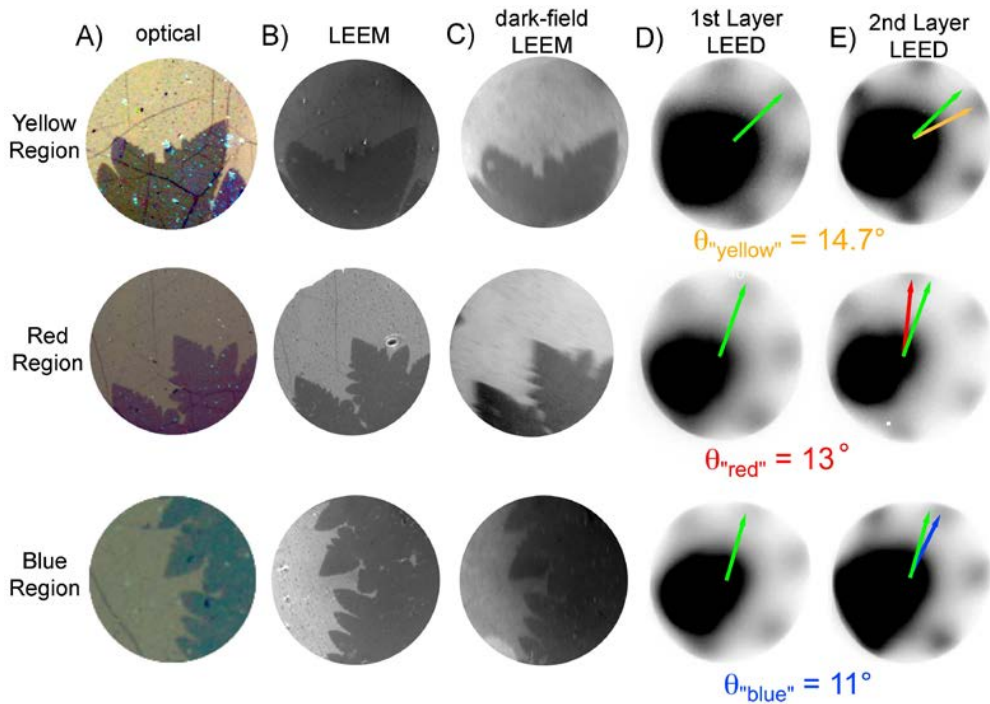

**Figure S3:** Comparison between the TBG domain color as seen optically and the twist angle θ as measured with LEED. Optical microscope images **(A)** of a yellow, red and blue domain. **(B)** LEEM images at the boundary between a single-layer and bilayer regions corresponding to the yellow, red and blue domains in **(A)**. **(C)** Dark-field LEEM image at the same regions in **(B)**, where an aperture was used to select one of the first order spots of the first graphene layer. **(D)** LEED images of the first graphene layer in **(A, B)**. **(E)** LEED pattern acquired over the "yellow", "red" and "blue" bilayer regions with the resulting twist angles. All optical and LEEM images have about 75 μm field of view.